\documentclass[twocolumn]{aastex63}
\usepackage{CJK}
\usepackage{amsmath}
\usepackage{xcolor}

\def\siggas{$\Sigma_{\rm{gas}}$}
\def\sigSFR{$\Sigma_{\rm{SFR}}$}
\def\sigM{$\Sigma_{\rm{M}}$}
\def\etaE{$\eta_E$}
\def\etam{$\eta_m$}
\def\etaZ{$\eta_Z$}
\def\etaEh{$\eta_{E,h}$}
\def\etamh{$\eta_{m,h}$}
\def\etaZh{$\eta_{{Z,h}}$}
\def\etaEc{$\eta_{E,c}$}

\def\es{$e_s$}
\def\esh{$e_{{s,h}}$}
\def\esc{$e_{{s,c}}$}
\def\fvh{$f_{{V,h}}$}
\def\msun{$M_\odot$}

\graphicspath{{./}{figures/}}

\begin{document}
\begin{CJK*}{UTF8}{gbsn}

\title{Simple Yet Powerful: Hot Galactic Outflows Driven by Supernovae}

\correspondingauthor{Miao Li }
\email{mli@flatironinstitute.org}

\author[0000-0003-0773-582X]{Miao Li   
(李邈)}
\affiliation{Center for Computational Astrophysics, Flatiron Institute, New York, NY 10010, USA}

\author[0000-0003-2630-9228]{Greg L. Bryan}
\affiliation{Center for Computational Astrophysics, Flatiron Institute, New York, NY 10010, USA}
\affiliation{Department of Astronomy, Columbia University, 550 West 120th Street, New York, NY 10027, USA}

\begin{abstract}
Supernovae (SNe) drive multiphase galactic outflows, impacting galaxy formation; however, cosmological simulations mostly use \textit{ad hoc} feedback models for outflows, making outflow-related predictions from first principles problematic. Recent small-box simulations resolve individual SNe remnants in the interstellar medium (ISM), naturally driving outflows and permitting a determination of the wind loading factors of energy \etaE, mass \etam, and metals \etaZ. In this Letter, we compile small-box results, and find consensus that the hot outflows are much more powerful than the cool outflows: (i) hot outflows generally dominate the energy flux, and (ii) their specific energy $e_{s,h}$ is 10-1000 times higher than cool outflows. Moreover, the properties of hot outflows are remarkably simple: $e_{s,h} \propto \eta_{E,h}/\eta_{m,h}$ is almost invariant over four orders of magnitude of star formation surface density. Also, we find tentatively that $\eta_{E,h}/\eta_{Z,h} \sim$ 0.5. If corroborated by more simulation data, these correlations reduce the three hot phase loading factors into one. Finally, this one parameter is closely related to whether the ISM has a ``breakout" condition. The narrow range of \esh\ indicates that hot outflows cannot escape dark matter halos with log $M_{\rm{halo}}\ [M_\odot] \gtrsim 12$. This mass is also where the galaxy mass-metallicity relation reaches its plateau, implying a deep connection between \textit{hot} outflows and galaxy formation. We argue that hot outflows should be included explicitly in cosmological simulations and (semi-)analytic modeling of galaxy formation.

\end{abstract}


\section{Introduction}
\label{sec:intro}

Supernovae (SNe) feedback is a critical ingredient in galaxy formation \citep[see recent reviews by][]{somerville15,naab17}. SNe drive turbulence into the ISM and launch galactic outflows, which are key to understanding the inefficiency of star formation \citep{mckee07} and the metal loss from galaxies \citep[e.g.][]{tremonti04,peeples14,maclow99}. 
However, feedback is arguably the least understood factor. Cosmological simulations, usually using \textit{ad hoc} models for SNe-driven feedback with fine-tuned parameters, cannot make feedback-related predictions from first principles. Galactic outflows are diffuse and multiphase, posing observational challenges\citep{veilleux05}; the evolution of multiphase gas is also an open question theoretically.

Recently, several groups have used 3D, kpc-scale simulations to investigate how SNe drive outflows from the ISM. These simulations are generally able to resolve the evolution of individual SNe remnants, which is critical to reaching numerical convergence and obtaining a multiphase ISM \citep[e.g.][]{kim15,hu19}. They model either a patch of the galaxy disk of massive systems like the Milky Way, or an isolated dwarf galaxy. The outflows, like the ISM, are multiphase, with typical temperature of $10^{6-7}$ K, $10^4$ K, and $10^2$ K, termed ``hot'', ``cool'', and ``cold'' phases. Different phases of outflows have very distinct properties, such as densities, velocities, and metallicities \citep[e.g.][]{creasey13,li17a,kim18,fielding18,hu19}. Therefore, the impact of different outflow phases on galaxy formation should be very different.

This Letter summarizes results from recent small-box simulations, highlighting that hot outflows are much more powerful than cool outflows. We thus advocate that hot outflows be employed explicitly and properly in cosmological simulations. Furthermore, we find that the hot outflows have simple scaling relations among their capacity to carry energy, mass and metals. 
The Letter is organized as follows: we describe the small-box simulations and define the loading factors in Section 2, present the results in Section 3, discuss the implications in Section 4, and summarize the findings and point to future work in Section 5.

\section{Data}

\begin{table}[]
\setlength\tabcolsep{0.6pt}
\caption{Data Source of Small-box Simulations}
\begin{tabular}{ccc} \hline
Reference        & h {[}kpc{]}\footnote{height above the midplane where the outflows properties are measured}     & Code       \\\hline
\cite{li17a}         & 1-2.5\footnote{outflow quantities are averaged over this range of height }& Enzo        \\
\cite{kim18},   & 1 & Athena MHD \\
 \& \textit{in prep}    & & \\
\cite{fielding18}  & 0.5            & Athena     \\
\cite{hu19}       & 1              & Gadget        \\
\cite{armillotta19}& 0.4            & GIZMO      \\
 \cite{emerick19}  &  1          &  Enzo  \\ 
\cite{martizzi16}   &  0.13, 0.15, 0.25\footnote{outflows are measured at different heights for different runs} & RAMSES     \\
\cite{creasey15}   & 0.5            & FLASH    \\ \hline
\end{tabular}
\end{table}

We focus on comparing different phases in the outflows from small-box simulations. Table 1 lists the data sources of the small-box simulations that are used in this Letter. The table includes all work to our knowledge that reported the loading factors of outflows in different phases. These investigations use many of the major (magneto-)hydrodynamic codes in the field. They have a length resolution of a parsec or a mass resolution of 10 \msun, generally capable of resolving the evolution of individual SNe remnants before reaching the cooling stage. This is necessary to avoid the overcooling problem, to generate a multiphase ISM and outflows, and to converge numerically \citep{simpson15,kim15,hu19}. This high resolution is currently not achievable in cosmological simulations, except for the smallest systems.

The outflows are measured above the midplane, at a height of 0.1-2.5 kpc, which is listed in Table 1. For each work, the measuring height is the same for different phases of outflows. 

Small-box simulations cover a wide range of gas surface density \siggas, star formation surface density \sigSFR, and mass surface density \sigM\ (which determines the gravitational potential). These conditions are listed in the extended version of Table 1\footnote{\url{https://github.com/limiao0611/loading-factors/wiki}}, where new data can be easily added when available.

The temperature division of hot and cool outflows in different work is slightly different, but typically at a few times $10^5$ K. The results are insensitive to the exact choice, since most mass is in well-divided temperature ranges, i.e., hot outflows are at $10^{6-7}$ K, and cool outflows around $10^4$ K. The cold phase in outflows is very minor, if existent at all, compared to the hot and cool phases \citep{li17a,hu19,kim18,armillotta19}. So we focus on the hot and cool phases in this Letter.

\subsection{Loading factors}

In small-box simulations, dimensionless ``loading factors" are used to quantify the loading efficiencies of SNe-driven outflows. The mass loading factor \etam\ and the energy loading factor \etaE\ are defined as follows: 
\begin{equation}
\eta_m \equiv \frac{\dot{M}_{\rm{out}}}{ \dot{M}_{\rm{SFR}}}, 
\label{eq:eta_m}
\end{equation}
\begin{equation}
\eta_E \equiv  \frac{\dot{E}_{\rm{out}}}{ \dot{E}_{\rm{SN}} } = \frac{\dot{E}_{\rm{out}}}{E_{\rm{SN}}  \dot{M}_{\rm{SFR}}/m_{*}}.
\label{eq:eta_E}
\end{equation}
where $\dot{M}_{\rm{out}}$, $\dot{E}_{\rm{out}}$ are the outflow rate of mass and energy (including both thermal and kinetic), which are measured from simulations; $\dot{M}_{\rm{SFR}}$ is the star formation rate,  $m_{*}$ is the mass of stars formed to produce one SN, and $E_{\rm{SN}}$ is the energy released by each SN.

The metal loading factor in the literature is defined in two different ways, which stems from the fact that small-box simulations usually only cover a short timescale compared to the Hubble time, and thus begin with a certain ISM metallicity $Z_{\rm{ISM}}$.
Therefore, metals in outflows come from two ``origins", that is, the metal mass outflow rate
\begin{equation}
    \dot{M}_{\rm{Z,out}} =  \dot{M}_{\rm{Z,out,SN}} + \dot{M}_{\rm{out}} \bar{Z}_{\rm{ISM}} , 
    \label{eq:M_Z_out}
\end{equation}
where the first term on the right $\dot{M}_{\rm{Z,out,SN}}$ is the metal outflow rate due to recent SNe enrichment, and the second term is the metal outflow rate due to $Z_{\rm{ISM}}$. As a result, one way to define the metal loading factor is
\begin{equation}
\eta'_{Z} \equiv \frac{\dot{M}_{\rm{Z,out}}}{\dot{M}_{\rm{Z,SN}}}
= \frac{\dot{M}_{\rm{Z,out}}}{m_{Z,\rm{SN}} \dot{M}_{\rm{SFR}} /m_* },
\label{eq:eta'_Z}
\end{equation}
where $\dot{M}_{\rm{Z,SN}}$ is the production rate of metals by SNe, and $m_{\rm{Z,SN}}$ is the metal mass released to the ISM per SN. This definition thus depends on $Z_{\rm{ISM}}$. The other way to define the metal loading factor is
\begin{equation}
    \eta_Z \equiv \frac{\dot{M}_{\rm{Z,out,SN}}}{ \dot{M}_{\rm{Z,SN}}   }.
\end{equation}
This is simply the fraction of metals produced by SNe that go into outflows, so is always $<$ 1 and is independent of $Z_{\rm{ISM}}$ adopted. The two metal-loading factors are connected through
\begin{equation}
\eta_Z  = \eta'_Z - \eta_m\bar{Z}_{\rm{ISM}}\frac{m_*}{m_{\rm{Z,SN}}}.
\label{eq:eta_Z_Z'}
\end{equation}
We use \etaZ\ in this Letter, and for work that reported $\eta'_Z$, we convert it into \etaZ\ using Eq. \ref{eq:eta_Z_Z'}. 

While the loading factors of hot outflows are almost independent of height, those of cool outflows decline quickly with increasing height within the small-box domain \citep{li17a,kim18,fielding18,hu19,armillotta19}. This is because the specific energy of the bulk cool outflows is comparable to the gravitational potential from the midplane to the measuring height ($\sim$ kpc); the specific energy of hot outflows, on the other hand, is much greater.

The three loading factors describe the mean properties of outflows. We use the reported loading factors from the small-box simulations. They are time-averaged values, and have an uncertainty of 10-20\% when extracted from the published figures. 

\section{Results}

\begin{figure}
\begin{center}
\includegraphics[width=0.50\textwidth]{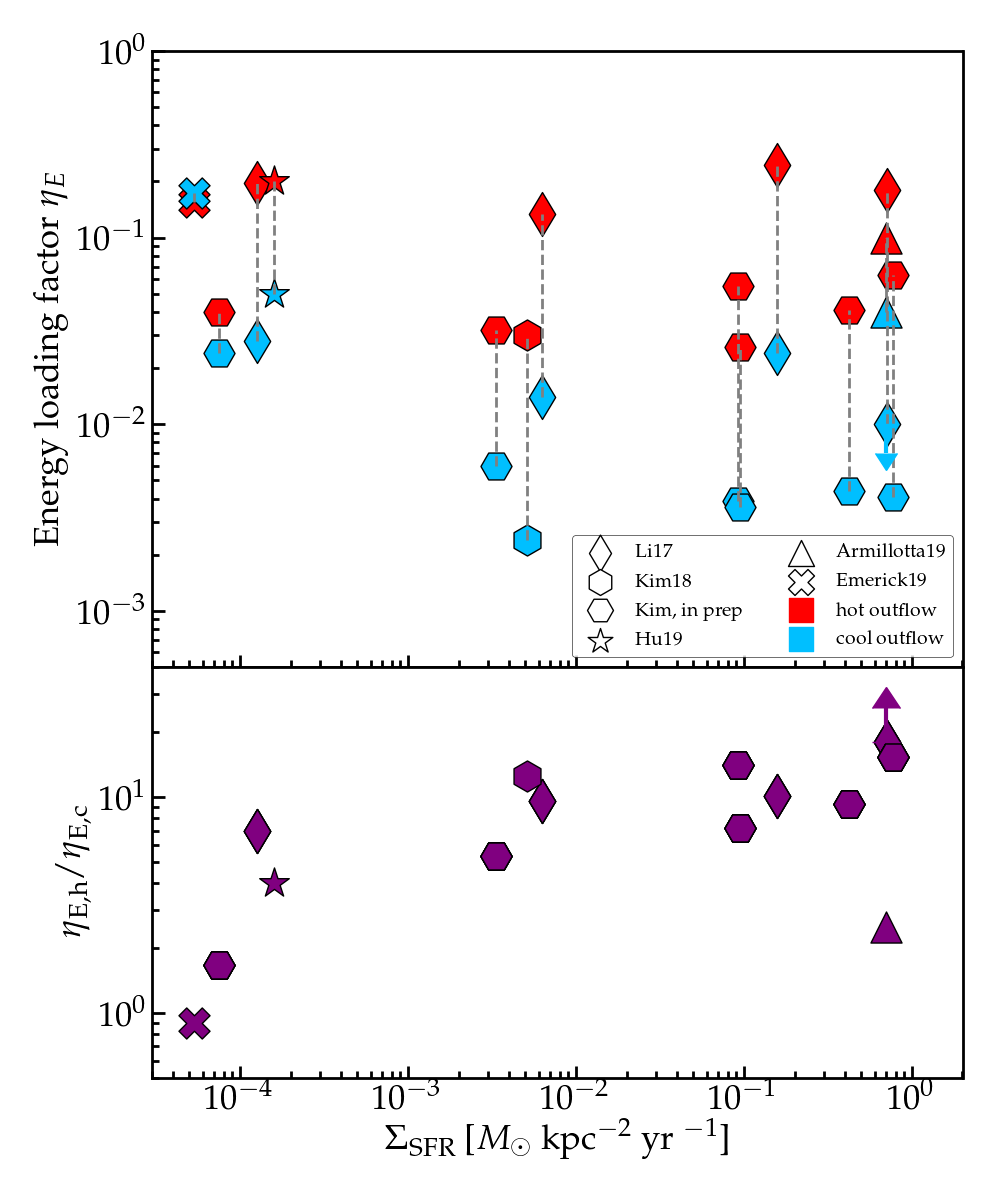}
\caption{Energy loading factor $\eta_E$ as a function of \sigSFR, for the hot (red) and cool (blue) outflows from small-box simulations. Each pair connected by a dash line is from the same run. The lower subplot shows the ratio of \etaE\ between the hot and cool outflows, which is 0.9--20. Hot outflows carry the majority of the outflow energy, except when \sigSFR\ is very low. 
}
\label{f:etaE}
\end{center}
\end{figure}

One important quantifier of the outflow power is $\eta_E$, the faction of SN energy that goes into outflows. Fig. \ref{f:etaE} shows $\eta_E$ as a function of \sigSFR\ from various small-box simulations. The red points indicate hot outflows and the blue points show cool outflows.
Each hot/cool pair connected by a dash line is from one simulation run (thus sharing the same \sigSFR).
Clearly, for every hot/cool pair, the hot outflows carry much more energy than the cool outflows, except one case which has the lowest \sigSFR \citep{emerick19}. The ratio of \etaE\ between the hot and the cool outflows \etaEh/\etaEc$=$0.9--20, as shown in the lower subplot. This indicates that hot outflows generally carry the majority of the outflow energy flux. Not shown in the plot are results from \cite{fielding18}, in which SNe are highly clustered in a molecular cloud and thus \sigSFR\ is not well-defined (it is scale-dependent). Nevertheless, their hot outflows have \etaEh$0.1-0.5$, and \etaEh/\etaEc$=$70-4000 for the four breakout cases (Fielding, private communication). This confirms that hot outflows dominate the energy partition (actually even more so) when star formation is highly concentrated.

One may notice that runs of similar \sigSFR\ can have \etaE\ different by a factor of up to 10. This can be because of the different gravitational potential adopted, and/or the scale height of SNe; \siggas\ is not a major factor since these simulations generally have their \siggas-\sigSFR\ along the Kennicutt-Schmidt relation \citep{kennicutt12}. But the main point we focus on here is that in all simulations, hot outflows carry most of the energy.

\begin{figure}
\begin{center}
\includegraphics[width=0.50\textwidth]{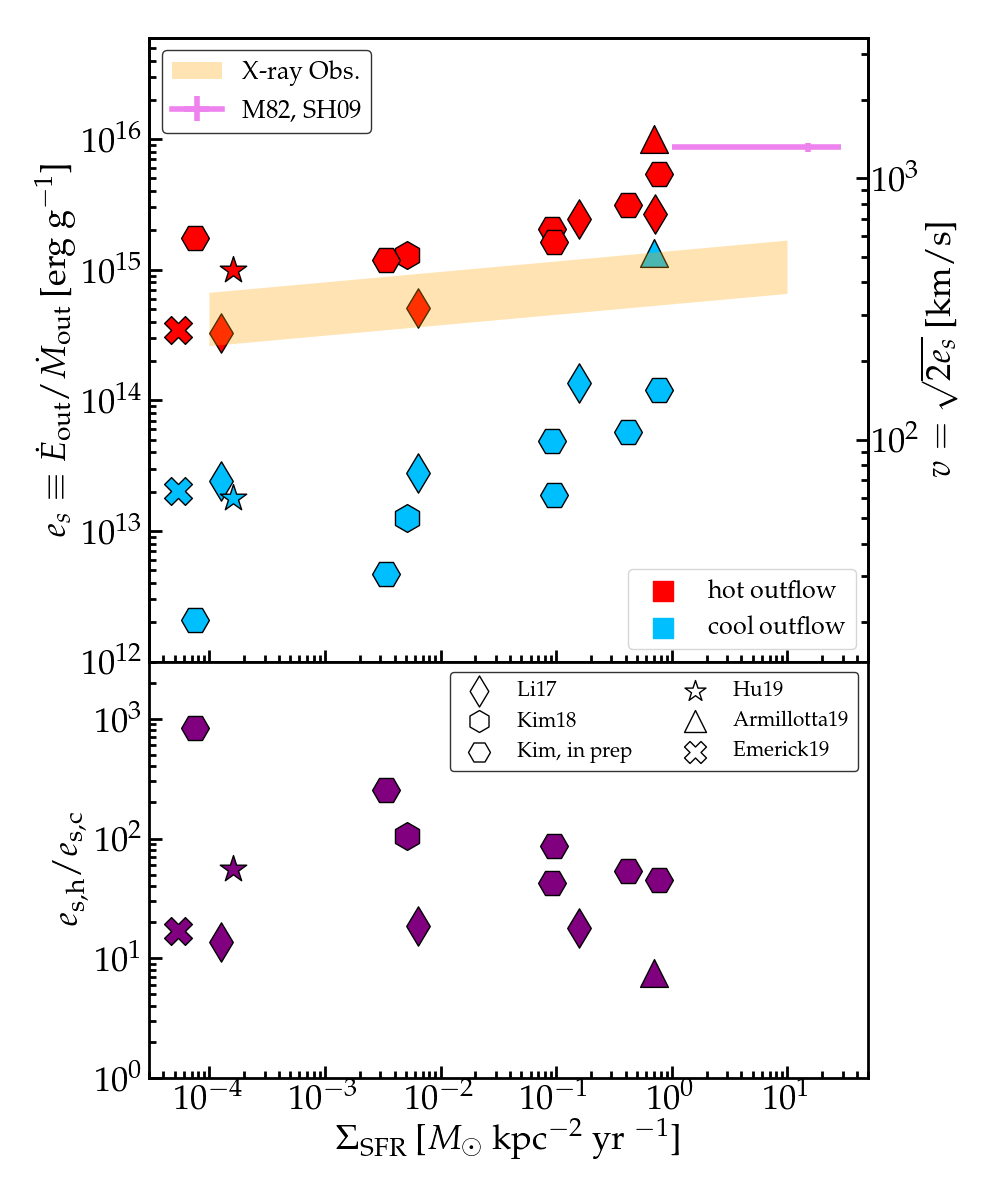}
\includegraphics[width=0.50\textwidth]{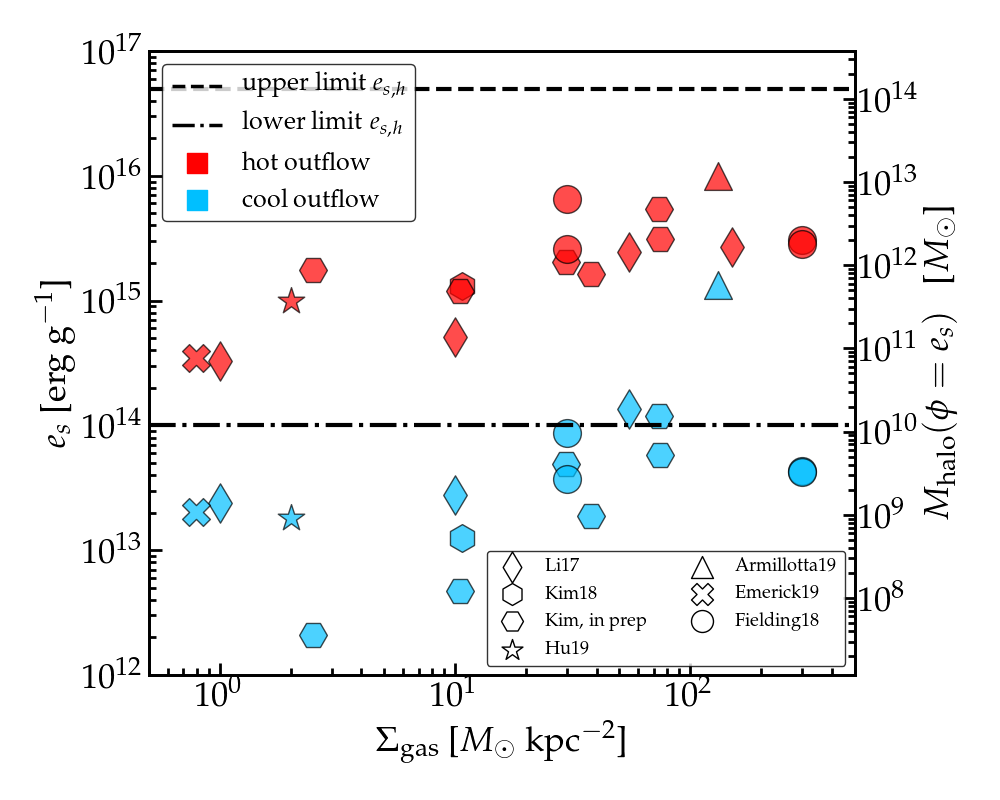}
\caption{Upper panel: specific energy \es\ of hot (red) and cool (blue) outflows as a function of \sigSFR. The right y-axis shows the terminal velocity $v\equiv \sqrt{2e_s}$. The orange region indicates the observed X-ray temperature of galactic coronae compiled by \cite{wang16}. The magenta cross is from the best-fit parameter for hot outflows in M82 \citep{strickland09}.  
The lower subplot shows the ratio $e_{s,h}/e_{s,c}$ for each small-box simulation. Hot outflows have \es\ greater than the cool ones by a factor of 10--1000. 
Lower panel: \esh\ as a function of \siggas. The dashed and dot-dashed lines bracket the possible range of \esh\ (see text for details). The actual \esh\ from small-box simulations have a narrower range. 
The right y-axis shows the dark matter halo mass with a potential $\phi = e_s$. Hot outflows have $e_s$ around the potential of 10$^{12\pm1}$ \msun\ halo.  
}
\label{f:es_gas}
\end{center}
\end{figure}

Another important parameter of outflows is the specific energy \es, defined as
\begin{equation}
e_s \equiv \frac{\dot{E}_{\rm{out}}}{\dot{M}_{\rm{out}}} =\frac{\eta_E E_{\rm{SN}}}{\eta_m m_*}.
\end{equation}
The upper panel of Fig. \ref{f:es_gas} shows \es\ as a function of \sigSFR, whereas the lower panel shows the same data as a function of \siggas, plus data from \cite{fielding18}. 
For a pair of hot/cool outflows from any simulation, \esh $\gg$ \esc. The ratio \esh/\esc\ is shown in the lower subplot, which ranges from 10-1000. This means that for a given gravitational potential, hot outflows can go much further than the cool outflows. The right y-axis of the upper panel shows the terminal velocity $v\equiv (2e_s)^{1/2}$.

The data also suggest that \esh\ increases with \sigSFR\ (and \siggas), but only weakly. A power-law fitting formula for \esh\ as a function of \sigSFR\ is
\begin{equation}
e_{\rm{s,h}} = 3.9\times 10^{15}\left(  \frac{\Sigma_{\rm{SFR}}  }{  \rm{M_\odot\ kpc^{-2}\ yr^{-1} }} \right)^{0.20\pm0.05}   \rm{erg}\ \rm{g}^{-1} . 
\end{equation}
Rewriting it in terms of the loading factors,
\begin{equation}
\begin{split}
\frac{\eta_{\rm{E,h}}}{\eta_{\rm{m,h}}}  =   0.78 & \left( \frac{ \Sigma_{\rm{SFR}} }{  \rm{M_\odot\ kpc^{-2}\ yr^{-1} }} \right)^{0.20\pm0.05} \times  \\ 
& \left( \frac{m_*}{100 \ M_\odot} \right)   \left( \frac{10^{51} \rm{erg}}{E_{\rm{SN}}} \right).
\end{split}
\label{eq:eta_e_m_ratio}
\end{equation}

The orange shaded area on the upper panel shows the observed X-ray temperature of galaxy coronae as a function of \sigSFR, $T_X=$ 0.56$\pm$0.13 keV $\Sigma_{\rm{SFR}}^{0.08\pm0.05}$, where \sigSFR\ is in unit $M_\odot$ kpc$^{-2}$ yr$^{-1}$ \citep{wang16}. The scaling relation indicates a weak (and positive) dependence of \esh\ on \sigSFR. 
Since $T_X$ includes only the thermal part of the specific energy, it is expected to be a lower limit to \esh. The small-box results are broadly consistent with the observational constraints.
The magenta error bar indicates the best-fit parameters for the hot outflows of M82, through detailed modelings of multiple metal lines and comparisons to X-ray observations \citep{strickland09} (with a relatively large uncertainty on \sigSFR). This is consistent with the extension of the small-box results.

Note that over 4 orders of magnitude of \sigSFR\ and 3 orders of magnitude of \siggas, \esh\ only varies within a factor of 30, log \esh\ [erg g$^{-1}$] = 14.5-16.
Admittedly, natural limits exist for \esh. If \es\ is too low the gas will not be ``hot'', so the lower limit of \esh\ is roughly $e_s(5\times 10^5$ K) $=10^{14}$ erg g$^{-1}$. The upper limit of \esh\ is that of SNe ejecta, i.e., $e_s \sim E_{\rm{SN}}/M_{\rm{SN}} \sim 10^{51} $ erg/10$M_\odot = 5\times 10^{16}$ erg g$^{-1}$.
The two limits are shown by the two black lines in the lower panel of Fig. \ref{f:es_gas}. Small-box results show that \esh\ has a narrower range than these limits. 
Radiative cooling and mass loaded from the ISM make the \esh\ smaller than the upper limit, whereas the presence of kinetic energy make \esh\ larger than the lower limit.

One can compare \es\ to the potential well of dark matter halos $\phi$ (from the galaxy to infinity), which gives simple estimates on whether or not outflows can escape from the halo. 
The right y-axis of the lower panel of Fig. \ref{f:es_gas} shows a mass of dark matter halo with $\phi =$ \esh\ on the left y-axis, i.e., $M_{\rm{halo}}(\phi =$\esh), where we use a simple scaling relation,
\begin{equation}
    \phi  = \frac{1}{2} (620\ \rm{km\ s^{-1}} )^2 \left( \frac{M_{\rm{halo}}}{10^{12} M_\odot} \right)^{2/3}.
\end{equation}
All the red points are around log $ M_{\rm{halo}}$ [\msun] $= 12 \pm 1$. Since hot outflows are heavily metal-enriched, whether they can leave the halo have important implications for metal loss in galaxy formation.
The observed mass-metallicity relation of galaxies reaches a plateau around a stellar mass of $10^{10.5} M_\odot$, corresponding to a halo mass of $10^{12}$ \msun \citep[e.g.][]{tremonti04,mannucci10,andrews13}.
This plateau may exist because metals cannot leave a system where hot outflows are gravitationally bound. More work is required to understand this in a cosmological setting, and to see how it compares to existing models for the mass-metallicity relation \citep[e.g.][]{ma16,torrey19}.

\begin{figure}
\begin{center}
\includegraphics[width=0.50\textwidth]{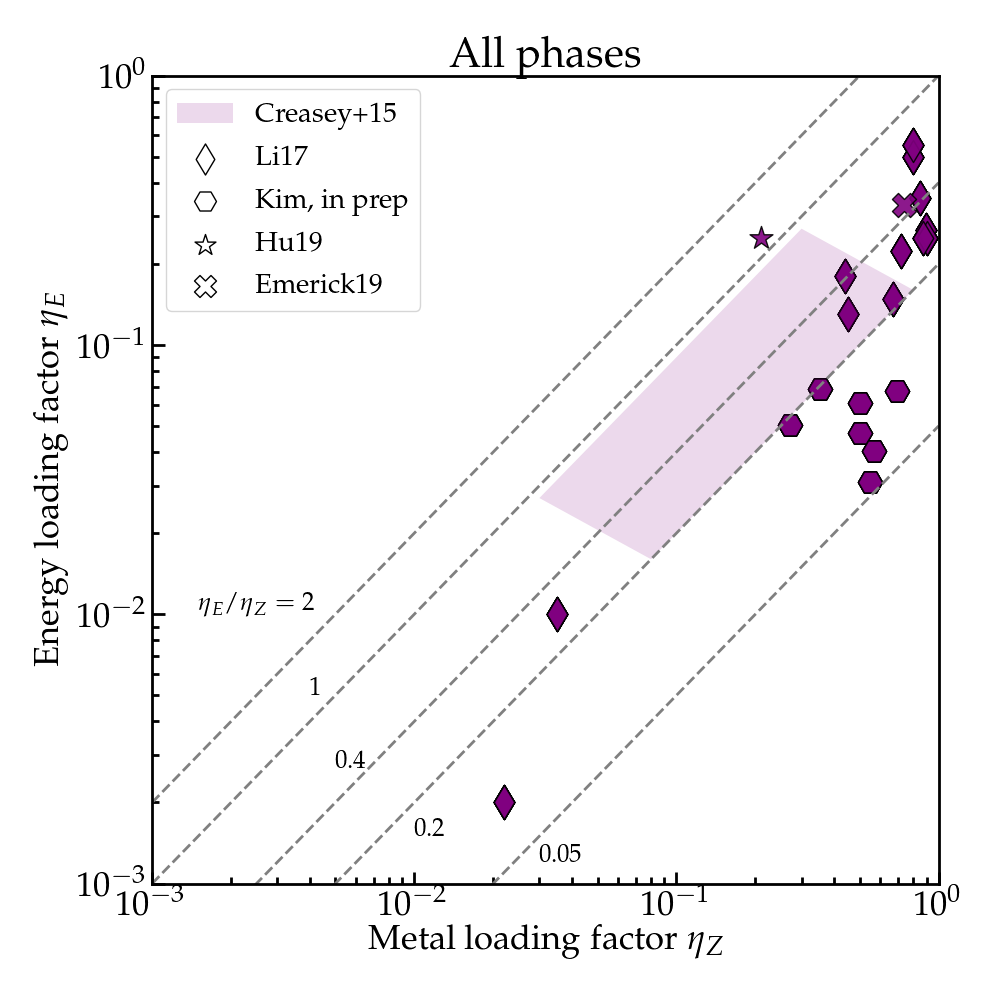}
\includegraphics[width=0.50\textwidth]{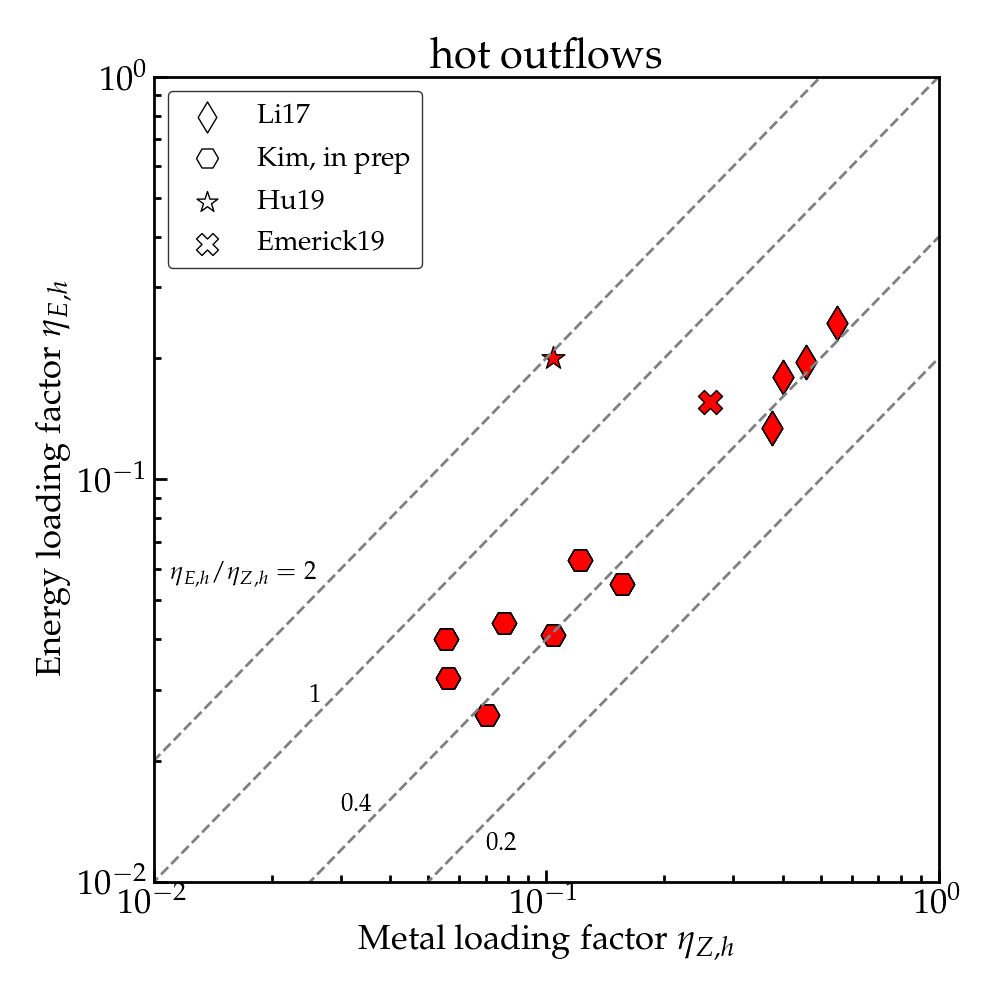}
\caption{\etaE\ versus \etaZ. The upper panel is for all phases, and the lower panel for hot outflows only. The dashed diagonal lines indicate constant $\eta_{E}/\eta_Z$, with values shown on the plots.
There is a positive correlation between \etaE\ and \etaZ, but the correlation is tighter when including hot outflows only (lower panel) . Some discrepancies exist. \cite{li17a}, \cite{emerick19}, and Kim et al. (\textit{in prep}) have \etaEh/\etaZh$\sim$ 0.5, whereas \cite{hu19} has $\sim$ 2. This may be due to the differences of numerical diffusion in grid- and particle-based codes. 
}
\label{f:etaE_etaZ}
\end{center}
\end{figure}

The energy and metal loading factors are positively correlated, as first pointed out and discussed in \cite{creasey15}. Their reported loading factors include all outflow phases, and the results (from a few dozen runs with varying \siggas, \sigSFR\ and \sigM) are represented by the shaded region in the upper panel of Fig. \ref{f:etaE_etaZ}. They found that \etaE/\etaZ $\sim$ 0.4. We add more data points from recent simulations, shown by the scattered points on the upper panel. While there is a general positive correlation between \etaE\ and \etaZ, the scatter is fairly large, with \etaE/\etaZ\ ranging from 0.05-1.2. 

We argue that a tighter correlation of \etaE\ and \etaZ\ should exist if we include hot outflows only. This is simply because SNe create and enrich hot phases simultaneously. As a result, metals, like energy, are preferentially retained in the hot phase. Indeed, hot outflows have a larger metallicity than the the cooler phases and that of the mean ISM \citep{maclow99,creasey15,li17a,hu19}. The lower panel of Fig. \ref{f:etaE_etaZ} show \etaEh\ versus \etaZh. Not many data are available, but interestingly enough, results from \cite{li17a}, \cite{emerick19}, and Kim et al. (\textit{in prep}) line up at 
\begin{equation}
   \frac{\eta_{\rm{E,h}}}{\eta_{Z,h}} \sim 0.48\pm0.11.
    \label{eq:eta_E_Z}
\end{equation}
Notably, \cite{hu19} has a much larger \etaEh/\etaZh$\sim$ 2. 
The exact reason for this discrepancy is unclear, but may be because there is no inter-particle diffusion in the Lagrangian code \cite{hu19} used; on the other hand, the grid-based codes used by \cite{li17a}, \cite{emerick19} and Kim et al. (\textit{in prep}) may have too much numerical diffusion. More data are needed to further examine the possible correlation of \etaEh\ and \etaZh, and a careful modeling of metal diffusion is essential to determine the precise scaling coefficient.

\begin{figure}
\begin{center}
\includegraphics[width=0.50\textwidth]{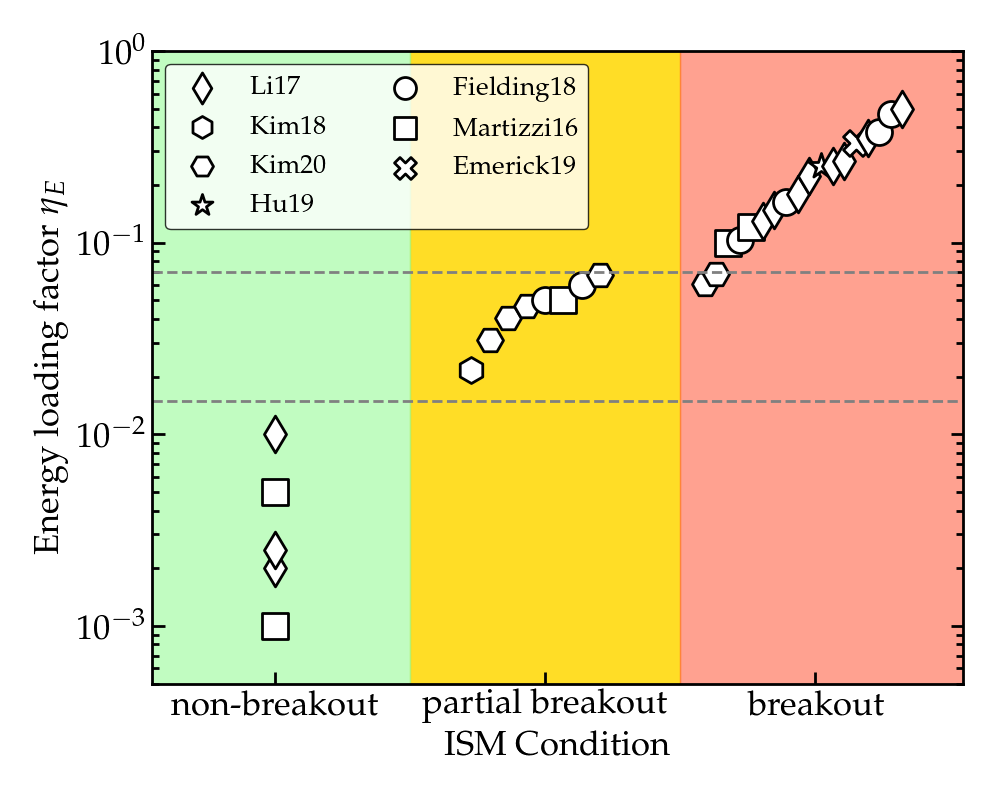}
\caption{Energy loading factors \etaE\ versus ISM condition. See the text for the quantitative definition of ``non-breakout'', ``partial breakout", and ``breakout" . Data are spread horizontally for the right two groups to avoid crowding. 
\etaE\ is consistently larger when ISM is in a breakout condition. The horizontal dashed lines mark the rough boundaries of \etaE\ for the three ISM conditions. }
\label{f:breakout}
\end{center}
\end{figure}

Finally, we show that the energy loading \etaE\ is closely related to whether a significant volume of the ISM is in the low-density, hot phase, i.e. ``breakout'' condition. Quantitatively, we define the breakout condition as when (a) the ISM has a volume-filling fraction of hot gas \fvh$\gtrsim$ 0.4, \textit{or} (b) the scale height of SNe is larger than that of the cool gas layer, ${h_{\rm{SN}}}/{h_{\rm{c,ISM}}} \gtrsim$ 1.2; $h_{\rm{c,ISM}}$ is measured after SNe have made the ISM multiphase and the cool ISM settles to a quasi-stable height. The non-breakout condition is defined as (a) \fvh$<$ 0.2,  \textit{and} (b) ${h_{\rm{SN}}}/{h_{\rm{c,ISM}}} < 0.8$. Conditions not satisfying ``breakout" or ``non-breakout" are categorized as ``partial breakout".
The values used are somewhat arbitrary, but the results are not sensitive to the exact choice; more data may help refine the exact criteria.

We collect all results where \etaE\ is reported and the ISM condition can be determined -- either from the reported \fvh\ or from slices where ${h_{\rm{c,ISM}}}$ can be measured. Note here we use \etaE\ instead of \etaEh\ to gain a larger sample (many results to date do not have phase-separate loading factors), with the knowledge that the hot outflows dominate the energy flux (Fig. \ref{f:etaE}). 
Fig. \ref{f:breakout} shows \etaE\ versus the ISM condition. To avoid crowding, we spread data horizontally for the right two categories (for no physical reason).
Overall, we find a very good correlation between \etaE\ and the ISM condition: \etaE$\gtrsim$ 0.08 when ISM is ``breakout"; \etaE$\lesssim$ 0.01 when ISM is ``non-breakout"; \etaE$\sim$ 0.01-0.08 when ISM is ``partial-breakout". We note that ``partial-breakout", in reality, can either be due to the final, quasi-steady ISM state falling into the criterion \citep{martizzi16}, or the ISM transitioning/cycling between the non-breakout and the breakout states \citep{kim18}. 

The correlation between \etaE\ and the ISM condition is not hard to understand: when the ISM has sufficient low-density ``holes", cooling is inefficient and energy from new SNe is easy to vent through. This has been confirmed observationally \citep{lopez11} and theoretically \citep[e.g.][]{deavillez04,li15,gatto17,steinwandel19}. 
Generally, the breakout condition is from a sufficiently large SNe density given a gas density. This is generally easier to achieve with a low gas density \citep[e.g.][]{li15}. 
In reality, effects that contribute to a breakout include:
clustered SNe forming superbubbles that break out of the ISM \citep[e.g.][]{maclow88,kim17,fielding18}, some SNe exploding at high latitude where gas densities are low \citep{li17a}, physical processes such as gravity and other stellar feedback creating an inhomogeneous medium \citep[e.g.][]{hopkins12,girichidis16}. Here we simply show that once a significant fraction of the ISM is in tenuous, hot condition, a significant fraction of SNe energy (10-60\%) can escape into the halo. This escaped energy can have important implications, such as blowing metals away, heating the circumgalactic medium (CGM)/intergalactic medium (IGM), and ultimately, regulating galaxy formation.

\section{Implications}

Hot outflows have a much larger specific energy than the cool phases at launching sites. Thus the relative importance of hot outflows and cool ones in different dark matter halos varies: cool outflows are more important in low-mass halos ($M\lesssim 10^{9-10} M_\odot$), since they can carry lots of mass and metals out of the halo \citep{hu19,emerick19}. In higher mass halos ($M\gtrsim 10^{10-11} M_\odot$), hot outflows would be much more important since most mass and metals that cool outflows carry will simply fall back to the galaxy. The way hot outflows impact galaxy formation likely differs from that of cool outflows. Hot outflows do not carry much mass, with \etamh $\sim 0.1--2$, which even decreases with increasing \sigSFR \citep{zhang14,li17a}. But they carry a significant fraction of SNe energy (unless SNe do not break out from the ISM) and are volume-filling. This means that hot outflows may help suppress cosmic accretion, especially the ``hot'' mode where gas comes in smoothly and spherically \citep{rees77,birnboim03,nelson03}. In other words, hot outflows act more as ``preventive" feedback, by suppressing gas from reaching galaxies in the first place, rather than ``ejective" feedback by propelling gas out of galaxies \citep{dave12}. 

Hot outflows carry lots of metals, which can travel far distances because of their large \esh. This can have important implications for the mass-metallicity relation of galaxies (see Section 3) and the existence of metals in the CGM and IGM \citep[e.g.][]{tumlinson17}. The impact of hot outflows driven by SNe should be studied in detail on large scales. 

Furthermore, the possible tight correlation between \etaEh\ and \etaZh\ can constrain \etaEh\, which is hard to infer observationally. Since the amount of ``missing metals'' from galaxies have been constrained observationally\citep[e.g.][]{peeples14,erb06},  then using Eq. \ref{eq:eta_E_Z}, one can estimate the amount of energy released from galaxies. This is potentially a useful way to constrain the power of energy feedback to the CGM/IGM.

Galaxy formation models, including cosmological simulations and (semi-)analytic modeling, depend sensitively on the parameters regarding galactic outflows \citep[e.g.][]{schaye15,pillepich18,forbes19}.
Current cosmological simulations usually use \textit{ad hoc} models for galactic outflows, though some ``zoom-in" simulations, e.g., FIRE, start to resolve a multiphase ISM and launch outflows more self-consistently \citep{hopkins12}. Outflows in cosmological simulations are predominantly in cold phase. A comprehensive comparison between outflows in small-box and cosmological simulations is not yet possible, since the latter generally reports outflows with global quantities like halo mass and star formation rate, but not with \sigSFR. That said, we compare the Milky Way case using Table 3 of \cite{muratov15}, which compiled outflow properties from a few large-scale simulations \citep{hopkins12,vogelsberger13,ford14}. The median launch velocities are a factor of a few larger than those of the cool outflows from small-box simulations, while the mass loading factors are broadly comparable. To enable a detailed and systematic comparison with small-box results, it would be very informative to correlate the multiphase outflow properties with local \sigSFR\ in large-scale simulations, for galaxies with different masses and star formation levels.

In light of the small-box results, we advocate incorporating the hot SNe-driven outflows explicitly and properly. This should be done separately from cool outflows, since mixing different phases can result in an unrealistically high cooling rate. The energy/mass/metal content of hot outflows can be added according to the local \sigSFR. One technical convenience for implementing hot outflows is that they do not require very high numerical resolution. It is a critical step to evaluate how the hot SNe-driven outflows transport metals and suppress cosmic inflows in a realistic cosmological context.

The hot outflows and their impact on the CGM and IGM are most directly probed by soft X-rays. Current observations provide important constraints about hot outflows/cosmic accretion, though generally limited to radii $\lesssim$ tens of kpc around massive galaxies. The situation can be greatly improved with the next generation of X-ray telescopes, such as Athena, Lynx and HUBS. The Sunyaev-Zel'dovich effect is another important way to constrain the hot CGM, the data of which will come in within years from low-noise CMB surveys using ACT, Simons Observatory, the South Pole Telescope, and others. In addition, hot outflows may partly become cooler phases as they propagate beyond galaxies, due to radiative cooling \citep{thompson16}, adiabatic expansion, and/or uplifting and interacting with pre-existing halo gas \citep{voit17}. The fate and impact of hot outflows should be investigated in much greater detail.

\section{Summary and Future work}

In this Letter, we summarize results from recent small-box simulations of SNe-driven outflows, with loading factors measured at $\sim$kpc above the launching sites. We find unanimous agreement that the hot outflows are much more powerful than the cooler phases. Furthermore, the hot outflows have intriguingly simple relations among the three loading factors. Specifically, 
\begin{enumerate}
    \item  The hot phase generally dominates the energy flux of outflows (Fig. 1).
    
    \item The specific energy of the hot outflows, $e_{\rm{s,h}}$, is 10-1000 times greater than the cool outflows, indicating that hot outflows can travel much further away from the galaxy (Fig 2.). 
    
    \item $e_{s,h}$ increases weakly with star formation rate surface density, $e_{s,h} \propto$ \sigSFR $^{0.2}$. 
    
    \item The narrow range of log $e_{s,h}$ [erg g$^{-1}$] $ =$ 14.5-16 indicates that hot outflows cannot escape DM halos with log $M_{\rm{halo}}/M_\odot \gtrsim 12\pm 1$. This implies a deep connection between the \textit{hot} outflows and the mass-metallicity relation of galaxies.

    \item Tentative results show that \etaEh\ and \etaZh\  have a linear correlation (Fig. 3, Eq. \ref{eq:eta_E_Z}). The correlations among loading factors, i.e., Eq. \ref{eq:eta_e_m_ratio} and \ref{eq:eta_E_Z}, indicate that for hot outflows, the three parameters describing the loading efficiencies can be reduced to one. 

    \item The one parameter of hot outflows is closely related to whether the ISM has a breakout condition (Fig. 4). 

\end{enumerate}

Clearly, more data from simulations with resolved ISM and SN remnants are necessary to robustly establish, or disapprove, the emerging correlations of hot outflows. Many such simulations already exist \citep[e.g.][]{girichidis16,marinacci19,vasiliev19}, so only computing the phase-specific loading factors is needed. 

There is still a long way to go before fully understanding how SNe drive multiphase outflows, as well as their impact on galaxy formation. Future work for small-box modelling includes, but is not limited to: investigating the discrepancies among simulations (e.g. Fig \ref{f:etaE_etaZ}); cover more the parameter space, e.g. extremely high \sigSFR\ (up to $10^{4}$ $M_\odot$ yr$^{-1}$ kpc$^{-2}$ \citep[e.g.][]{heckman16}; examine additional physics, such as cosmic rays \citep{simpson16}; study the interaction among different gas phases \citep[e.g.][]{gronke19}. Future cosmological simulations with well-established, physically-based feedback models are essential to a predictive galaxy formation theory.

\vspace{0.2in}

\section*{Acknowledgement}
We thank the referee for helpful comments.
We thank Chang-Goo Kim and Eve Ostriker for sharing their simulation results and helpful discussions, and Drummond Fielding and Andrew Emerick for calculating quantities used in this paper. GB acknowledges financial support from NSF (grant AST-1615955, OAC-1835509), and NASA (grant NNX15AB20G), and computing support from NSF XSEDE.


\end{CJK*}

\end{document}